\documentclass[11pt]{article}
\usepackage[margin=2.6cm]{geometry}
\usepackage[T1]{fontenc} 
\usepackage[english]{babel} 
\usepackage[utf8]{inputenc}
\usepackage{indentfirst}
\usepackage{amsmath}
\usepackage{amssymb}
\usepackage{url}
\usepackage{graphicx}
\usepackage{braket}
\usepackage{slashed}
\usepackage{cancel}
\usepackage{multirow}
\usepackage[font=footnotesize,labelfont=bf]{caption}
\usepackage[usenames, dvipsnames]{xcolor}
\usepackage{booktabs}
\usepackage{dsfont}
\usepackage{soul}
\usepackage{pdflscape}
\usepackage{float}

\newcommand{\CP}{$C\!P$}
\newcommand{\LCPT}{L$N_c$ $\chi$PT}
\newcommand{\SLV}[1]{\boldsymbol{#1}}

\usepackage[colorlinks=true, urlcolor=MidnightBlue, linkcolor=MidnightBlue, 
citecolor=Mahogany, pdfborder={0 0 0}]{hyperref}
\usepackage[sort&compress,numbers]{natbib}
\usepackage{doi}
\usepackage[capitalise]{cleveref}

\title{\Large\bfseries{New-physics signatures via $C\!P$ violation in $\eta^{(\prime)}\to\pi^0\mu^+\mu^-$ and $\eta^\prime\to\eta\mu^+\mu^-$ decays}}
\author{\normalsize{
        Rafel Escribano{\color{Mahogany}\thanks{rescriba@ifae.es}}$^{\color{Mahogany}{\ a,b}}$, 
        Emilio Royo{\color{Mahogany}\thanks{eroyo@ifae.es}}$^{\color{Mahogany}{\ a,b}}$,      
        Pablo Sanchez-Puertas{\color{Mahogany}\thanks{psanchez@ifae.es}}$^{\color{Mahogany}{\ b}}$
        }\vspace{0.2cm}\\
        {\small{$^{\color{Mahogany}{a}}$\textit{Grup de F{\'isica} Te{\`o}rica,
        Departament de F{\'i}sica,}}}\\
        {\small{\textit{Universitat Aut{\`o}noma de Barcelona, E-08193 Bellaterra (Barcelona), Spain}}}\\
        {\small{$^{\color{Mahogany}{b}}$\textit{Institut de F{\'i}sica d'Altes Energies (IFAE)
        and
        Barcelona Institute of Science and Technology (BIST),}}}\\
        {\small{\textit{Campus UAB, E-08193 Bellaterra (Barcelona), Spain}}}
}
\date{}

\begin{document}
\renewcommand{\abstractname}{\vspace{-\baselineskip}} 
\maketitle


\begin{abstract}

In this work we investigate the prospect of observing new-physics signatures via $C\!P$ violation in $\eta^{(\prime)}\to\pi^0\mu^+\mu^-$ and $\eta^\prime\to\eta\mu^+\mu^-$ decays at the REDTOP experiment. We make use of the SMEFT to parametrise the new-physics $C\!P$-violating effects and find that the projected REDTOP statistics  are not competitive with respect to nEDM experiments. This reasserts the $\eta\to\mu^+\mu^-$ process as the most promising channel to find $C\!P$-violation at this experimental facility.

\end{abstract}


\section{Introduction}
\label{sec:intro}

Over the past few decades, high-energy particle colliders have not succeeded in the quest for finding evidence for physics beyond the Standard Model (BSM). The purpose of large experiments such as the Large Hadron Collider (LHC), apart from settling down the question around the source of electroweak symmetry breaking, was to provide experimental evidence for either supersymmetric particles or extra dimensions or both, as they enjoy from strong theoretical motivation based on naturalness arguments, but this has not happened.

The current lack of experimental evidence for new physics in direct searches, that would help guide theoretical effort, is forcing the community to increase their focus on low-energy, high luminosity precision measurements that attempt to find effects from BSM physics by looking for small discrepancies between SM predictions and measurements. To this end, one focuses on processes whose SM contribution is very precisely known or that have a very small SM background, hence, any positive experimental finding would be a confirmation for new physics. Accordingly, interest in BSM searches in meson factories has significantly increased in recent years,\footnote{Another example of a low-energy experiment that has recently attracted much interest is the Muon g-2 at Fermilab~\cite{Abi:2021gix}.} as they can very precisely measure branching ratios of rare decays and test for violations of the basic symmetries. As an example, the observation of $C\!P$ violation in processes mediated by the strong or electromagnetic interactions would be an unambiguous sign of new physics and the study of the $\eta$ and $\eta^\prime$ decays represents the perfect laboratory for this endeavour. This is because both mesons are eigenstates of the $C$, $P$, $C\!P$ and $G$ operators (i.e.~$I^GJ^{PC}=0^+0^{-+}$) and their additive quantum numbers are zero, which amounts to all their decays being flavour conserving. 
As a consequence, and unlike flavoured meson decays, they can be used to test $C$ and $C\!P$ symmetries, provided a large sample of $\eta$ and $\eta^{\prime}$ mesons is available. Furthermore, their strong and electromagnetic decays are forbidden at lowest order, increasing their sensitivity to rare decays. 

In this context, a new experiment named REDTOP has been proposed~\cite{Gatto:2016rae,Gatto:2019dhj}, which aims at producing the largest sample of $\eta$ and $\eta'$ mesons envisaged thus far, and is considering implementing dedicated detectors to perform muon polarimetry. In order to set their priorities, it is crucial to assess the most promising channels and the physics within reach. 
In Ref.~\cite{Sanchez-Puertas:2018tnp}, the possibility of observing new physics signatures via $C\!P$-violating effects at REDTOP was assessed using muon polarisation observables in $\eta$ leptonic decays. In particular, the purely leptonic channels $\mu^+\mu^-$, $\mu^+\mu^-\gamma$, and $\mu^+\mu^-\ell^+\ell^-$ were studied, finding that $C\!P$ violation in the $\mu^+\mu^-$ final state could be observed at REDTOP, while evading neutron electric dipole moment (nEDM) constraints.

In the present work, we investigate the suitability of some $\eta$ and $\eta^\prime$ semileptonic decays, which were not covered in the previous study as they require a dedicated analysis of hadronic matrix elements. In particular, we investigate the $\eta^{(\prime)}\to\pi^0\mu^+\mu^-$ and $\eta'\to\eta\mu^+\mu^-$ decays using the SM effective field theory (SMEFT) as the general framework to capture new physics.\footnote{An investigation of the $\eta^{(\prime)}\to\pi^+\pi^-\mu^+\mu^-$ decays is also underway \cite{Maximilian}.} 
Using muon polarisation observables, we quantify the sensitivity that could be achieved at REDTOP for the relevant $C\!P$-violating Wilson coefficients. Our results show that these decays are not competitive when confronted against the stringent bounds derived from nEDM and $D_s^-\to \mu\bar{\nu}_{\mu}$ decays. This contrasts with the $\eta\to\mu^+\mu^-$ decay that evades these bounds and ought to receive the highest priority. 

The article is structured as follows. In \cref{decayamp}, we discuss the general properties of the decay amplitudes and narrow down the range of SMEFT operators that are relevant to our study. In \cref{app:hadmatrix}, we present the theoretical expressions for the required hadronic matrix elements obtained using large-$N_c$ chiral perturbation theory (\LCPT{}). The polarised decay widths and the asymmetries that quantify the $C\!P$-violating effects in $\eta^{(\prime)}\to\pi^0\mu^+\mu^-$ and $\eta'\to\eta\mu^+\mu^-$ decays are analysed in \cref{poldecasym}. The results from our investigation are presented in \cref{resultsdisc} and we briefly discuss their implications. Finally, in \cref{conclusions} we provide a summary of the work carried out and some final conclusions.
\\


\section{Decay amplitudes}\label{decayamp}

Defining the following momenta $q=p_{\mu^+} +p_{\mu^-} = p_{\eta^{(\prime)}}+p_{\pi(\eta)}$, $\bar{q}= p_{\mu^+} -p_{\mu^-}$, and $k=p_{\eta^{(\prime)}}-p_{\pi(\eta)}$, the most general form factor decomposition for $\braket{\mu^+\mu^- | i T | \eta^{(\prime)}\pi^0(\eta)} = i\mathcal{M}(2\pi)^4\delta(p_{\mu^+} +p_{\mu^-} -p_{\eta^{(\prime)}}-p_{\pi(\eta)})$ is
\begin{equation}\label{amplitude}
\mathcal{M} = m_{\mu}(\bar{u}v) F_1 +(\bar{u}i\gamma^5v) F_2 +(\bar{u}\slashed{k}v) F_3  + i(\bar{u}\slashed{k}\gamma^5v) F_4 \ ,
\end{equation}
where the $F_i\equiv F_i(q^2,\bar{q}\cdot k)$ form factors have been introduced. The connection to the $\eta^{(\prime)}\to\pi^0\mu^+\mu^-$ and $\eta^{\prime}\to\eta\mu^+\mu^-$ decays is obtained via crossing symmetry with $k\to p_{\eta^{(\prime)}}+p_{\pi(\eta)}$. 
General considerations on discrete symmetries can be used to show that electromagnetic interactions can only contribute to the $F_1(q^2,[\bar{q}\cdot k]^{2n})$ and $F_3(q^2,[\bar{q}\cdot k]^{2n+1})$ form factors, with $n=0,1,2\ldots$, 
and that they can, in turn, be expressed in terms of the $\Sigma$ and $\Omega$ parameters from Ref.~\cite{Escribano:2020rfs} as $F_1=\Sigma$ and $F_3 =\frac{1}{2}\Omega$. Furthermore, tree-level electroweak contributions appear via intermediate Higgs-boson exchange only, which contribute to the $C$- and $P$-conserving $F_1$ form factor, providing an unimportant correction to the present study. At higher orders, electroweak contributions to $F_{2,4}(q^2,[\bar{q}\cdot k]^{2n+1})$ of $C$- and $P$-odd nature can appear via $\gamma Z$ boxes, but these are $C\!P$-even and are, once again, irrelevant to the observables in this study.

Turning to the BSM $C\!P$-violating contribution, which requires a careful study of the underlying hadron dynamics, one starts by assuming that the SMEFT provides a correct description of Nature. Accordingly, new physics degrees of freedom are expected to lie above the electroweak scale and, therefore, only the SM particle spectra are considered. In addition, the new-physics effects come from higher dimension operators, that are suppressed by increasing powers of a large energy scale, starting with $D = 6$ so long as $B$-$L$ number conservation is assumed. 
In particular, the contribution from the different operators were outlined in Ref.~\cite{Sanchez-Puertas:2018tnp} and we briefly recapitulate here. Quark and lepton EDM operators are highly constrained by nEDM bounds; likewise, $C\!P$ violation in the hadronic sector requires $C\!P$-violating form factors with an additional electromagnetic $\alpha$ suppression, required in order to couple hadrons to leptons, which renders any such contribution not competitive. In addition, vector, axial and tensor operators have a vanishing coupling to the $\eta^{(\prime)}\pi^0$ and $\eta^{\prime}\eta$ systems based on discrete symmetries. Finally, Fermi operators involving quarks and leptons provide the most significant contribution and, thus, the operators that are considered in this study are\footnote{Interestingly, these operators generate the desired $C\!P$-odd contribution to our processes at tree level whilst, for the nEDM, contributions appear at the two-loops order weakening the nEDM bounds.} 
\begin{equation}
\label{eq:operators}
\mathcal{O}_{\ell edq}^{prst} = (\bar{\ell}^i_p e_r)(\bar{d}_s q^i_t) \ , \qquad
\mathcal{O}_{\ell equ}^{(1)prst} = (\bar{\ell}^i_p e_r)(\bar{q}^j_s u_t)\epsilon_{ij} \ ,
\end{equation}
where $pr\!st$ are family indices (i.e.~$p,r,s,t=1$ or $2$) \cite{Grzadkowski:2010es}. These operators produce a non-vanishing $C\!P$-odd $F_2$ form factor\footnote{Since our focus is on $C\!P$-violating effects, we are only concerned with the corresponding imaginary parts, as in Ref.~\cite{Sanchez-Puertas:2018tnp}.}
\begin{align}
\nonumber
\label{eq:F2smeft}
F_2 = &
\left[\operatorname{Im} c_{\ell edq}^{2211} \bra{0} \bar{d}d \ket{\eta^{(\prime)}\pi^0(\eta)} +\operatorname{Im} c_{\ell edq}^{2222} \bra{0} \bar{s}s \ket{\eta^{(\prime)}\pi^0(\eta)} + \right.\\
& \left. - \operatorname{Im} c_{\ell equ}^{(1)2211} \bra{0} \bar{u}u \ket{\eta^{(\prime)}\pi^0(\eta)} \right]\! /v^{2} \ ,
\end{align}
where $v^2 = 1/(\sqrt{2}G_F)$ and the corresponding hadronic matrix elements need a careful treatment that we discuss in the following section within the framework of \LCPT{}. To conclude this section, it is worth highlighting that at this order in the SMEFT there is no contribution to $F_4$.
\\


\section{Hadronic matrix elements}
\label{app:hadmatrix}

The matrix elements of the scalar currents (cf.~\cref{eq:F2smeft}) required for the calculation of the longitudinal and transverse asymmetries (cf.~Eqs.~(\ref{eq:AL}) and (\ref{eq:AT})) can be calculated within the framework of \LCPT{}, see Refs.~\cite{Kaiser:2000gs,Herrera-Siklody:1996tqr,Herrera-Siklody:1997pgy,Guo:2015xva,Bickert:2016fgy}. In the following, we evaluate them at NLO, after renormalising the fields and diagonalising the mass matrix (see, e.g., Appendix~B in Ref.~\cite{Escribano:2010wt} for a detailed account of the procedure). To simplify the expressions, we adopt the approach from Ref.~\cite{Escribano:2016ntp}, assuming that the $q^2$ dependence of the associated form factors is saturated by the corresponding scalar resonances, and make use of the resonance chiral theory (R$\chi$T) prediction $4L_5/F_0^2=8L_8/F_0^2=1/M_S^2$ from Ref.~\cite{Ecker:1988te,Ecker:1989yg,Cirigliano:2006hb}. Furthermore, to obtain non-vanishing $\bra{0}\bar{s}s\ket{\pi^0\eta^{(\prime)}}$ matrix elements, one needs to take into account isospin-breaking effects.\footnote{This is particularly important in this study since the Wilson coefficients associated to the strange quark are comparatively far less constrained by the nEDM bounds than the light-quark ones, see Ref.~\cite{Sanchez-Puertas:2018tnp}.} To this end, we follow the procedure from Ref.~\cite{Escribano:2020jdy} keeping only the leading isospin-breaking terms. Our results for the $\eta\pi^0$ matrix elements are 
\begin{align}
 \nonumber
 \bra{0}\bar{u}u/\bar{d}d\ket{\eta\pi^0} &{}= 
 \pm B_0\left[\left(1-\frac{m_\eta^2-m_\pi^2}{M_{S}^2}\right) \left(\cos{\phi_{23}} \pm\epsilon_{13}\sin{\phi_{23}}\right) + \right. \\ 
 & \quad \left. - \left(\cos{\phi_{23}}-\frac{\sin{\phi_{23}}}{\sqrt{2}}\right)\frac{\tilde{\Lambda}}{3}\right]
 \left(\frac{M_S^2}{M_S^2-s}\right) \ ,  \\[1 ex] \nonumber
   \bra{0}\bar{s}s\ket{\eta\pi^0} &{}= -2B_0\epsilon_{13}\left[\left(1-\frac{m_\eta^2+3m_\pi^2-4m_{K}^2}{M_S^2}\right)\sin{\phi_{23}} \right. + \\ 
  & \quad + \left. \frac{\tilde{\Lambda}}{3}\left( \frac{\cos{\phi_{23}}}{\sqrt{2}} - \sin{\phi_{23}} -\frac{\epsilon_{12}\sin{\phi_{23}}}{\sqrt{2} \epsilon_{13}} \right) \right]
   \left(\frac{M_S^2}{M_S^2-s}\right) \ , 
\end{align}
where we have introduced the scale invariant parameter $\tilde{\Lambda} = \Lambda_1 -2\Lambda_2$, $\phi_{23}$ is the $\eta$-$\eta^\prime$ mixing angle in the quark-flavour basis, $\epsilon_{12}$ and $\epsilon_{13}$ are first order approximations to the corresponding $\phi_{12}$ and $\phi_{13}$ isospin-breaking mixing angles in the $\pi^0$-$\eta$ and $\pi^0$-$\eta^\prime$ sectors, respectively (see Ref.~\cite{Escribano:2020jdy} for further details), and $M_S$ is the mass of a generic octet scalar resonance. 
The corresponding expressions for $\eta\to\eta'$ can be obtained by substituting $\cos{\phi_{23}} \to \sin{\phi_{23}}$, $\sin{\phi_{23}} \to -\cos{\phi_{23}}$ and $m_{\eta} \to m_{\eta'}$. For the $\eta'\to\eta\mu^+\mu^-$ decay, the matrix elements read
\begin{align}
 \nonumber
 \bra{0}\bar{u}u/\bar{d}d\ket{\eta^\prime\eta} &{}= 
 B_0\left[\left(1-\frac{m_{\eta^\prime}^2+m_\eta^2-2m_\pi^2}{M_S^2}\right)\left(\frac{ \sin{2\phi_{23}}}{2} \mp\epsilon_{13} \cos{2\phi_{23}}\right) + \right.  \\
 & \quad \left. -\left(\frac{\cos{2\phi_{23}}}{\sqrt{2}} +\sin{2\phi_{23}}\right)\frac{\tilde{\Lambda}}{3}
 \right]\left(\frac{M_S^2}{M_S^2-s}\right) \ , \\[1 ex] \nonumber
 \bra{0}\bar{s}s\ket{\eta^\prime\eta} &{}= -B_0\left[\left(1-\frac{m_{\eta^\prime}^2+m_\eta^2+2m_\pi^2-4m_{K}^2}{M_S^2}\right)\sin{2\phi_{23}} + \right. \\
& \quad \left. +\left(\sqrt{2} \cos{2\phi_{23}} - \sin{2\phi_{23}}\right)\frac{\tilde{\Lambda}}{3} 
\right]\left(\frac{M_S^2}{M_S^2-s}\right) \ . 
\end{align}
The numerical values that we employ are $B_0=m_{\pi^0}^2/2\overline{m}=2.64^{+0.11}_{-0.42}$ GeV at a renormalisation scale of $\mu = 2$ GeV, $M_S=980$ MeV \cite{Zyla:2020zbs}, $\tilde{\Lambda} = -0.46\pm0.19$ from lattice QCD~\cite{Bali:2021qem}, which is in agreement with other phenomenological results~\cite{Leutwyler:1997yr,Guo:2015xva,Bickert:2016fgy,Benayoun:1999au}, and the mixing parameters $\phi_{23}=(41.5\pm0.5)^{\circ}$, $\epsilon_{12}=(2.4\pm1.0)\%$ and $\epsilon_{13}=(2.5\pm 0.9)\%$ from Ref.~\cite{Escribano:2020jdy}. For the other masses we also take the PDG values.

Before concluding this section, two remarks are in order: first, the matrix elements of the strange quark scalar current with $\eta^{(\prime)}\pi^0$ are suppressed by the isospin symmetry-breaking parameter $\epsilon_{13}$ and, second, the contribution of the $\tilde{\Lambda}$ parameter is in general significant for the matrix elements involving the $\eta'$. 
\\


\section{Polarised decays and asymmetries}\label{poldecasym}

Let us now compute the squared amplitude from \cref{amplitude}, $|\mathcal{M}(\lambda\SLV{n},\bar{\lambda}\bar{\SLV{n}})|^2$, for the polarised decays that we are investigating. Using the conventions for the kinematics and the phase space given in \cref{app:kin}, and neglecting any contribution from the $F_4$ form factor as already mentioned at the end of \cref{decayamp}, we find
\begin{align}
\nonumber
\label{eq:masterf}
|\mathcal{M} (\lambda\SLV{n},\bar{\lambda}\bar{\SLV{n}}) |^2 &= \frac{1}{4}\Big[
      c_1|F_1|^2 +c_2|F_2|^2  + c_3|F_3|^2 +c_{13}^R \operatorname{Re}F_1F_3^* +c_{13}^I \operatorname{Im}F_1F_3^* \\   
 & \ \ \ +c_{12}^R \operatorname{Re}F_1F_2^* +c_{12}^I \operatorname{Im}F_1F_2^*      
    +c_{23}^R \operatorname{Re}F_2F_3^* +c_{23}^I \operatorname{Im}F_2F_3^* \Big] \ ,
\end{align}
where $\SLV{n}(\bar{\SLV{n}})$ is the $\mu^+(\mu^-)$ spin-polarisation axis defined in the $\mu^{\pm}$ rest frames and $\lambda=\pm$ denotes the two spin states. Note that terms in the first and second lines 
are $C\!P$ conserving and violating, respectively, provided that $F_{1,2}\equiv F_{1,2}(s,[\bar{q}\cdot k]^{2n})$ and $F_{3}\equiv F_{3}(s,[\bar{q}\cdot k]^{2n+1})$. The coefficients in \cref{eq:masterf} are given in \cref{app:polamps} and are the necessary input for implementation in the {\textsc{Geant4}} software~\cite{Agostinelli:2002hh}. The polarisation of the muons, however, cannot be directly measured and must be inferred from the velocities of the $e^{\pm}$ associated to the corresponding $\mu^{\pm}$ decays. Using the expressions provided in \cref{app:polamps} and making use of the spin-density formalism~\cite{Haber:1994pe}, one finds
\begin{align}
\nonumber
\label{eq:difdec}
d\Gamma =& \frac{ds dc\theta}{64(2\pi)^3} \frac{\lambda^{1/2}_K \beta_{\mu}}{m_{\eta^{(\prime)}}^3} \!
\left[\frac{d\Omega}{4\pi}dx \ n(x)\right]\! \left[\frac{d\bar{\Omega}}{4\pi}d\bar{x} \ n(\bar{x})\right] \!
\Big[
   \tilde{c}_1|F_1|^2 + \tilde{c}_3|F_3|^2 \\
   &+\tilde{c}_{13}^R \operatorname{Re}F_1F_3^*+\tilde{c}_{13}^I \operatorname{Im}F_1F_3^* +\tilde{c}_2|F_2|^2 +\tilde{c}_{12}^R \operatorname{Re}F_1F_2^* +\tilde{c}_{12}^I \operatorname{Im}F_1F_2^* \nonumber \\ 
   &+ \tilde{c}_{23}^R \operatorname{Re}F_2F_3^* +\tilde{c}_{23}^I \operatorname{Im}F_2F_3^* \Big] \ ,
\end{align}
where the 3-body phase-space description from \cref{app:kin} has been employed for the initial $\eta^{(\prime)}\to\pi^0\mu^+\mu^-$ and $\eta^{\prime}\to\eta\mu^+\mu^-$ decays, and the first two brackets 
account for the 
phase space of the subsequent $\mu^{\pm}$ decays, cf.~\cref{app:polmudec}. The coefficients in \cref{eq:difdec} are calculated from those in \cref{app:polamps} and read
\begin{align}
\tilde{c}_1 ={}& 2\beta_{\mu}^2 m_{\mu}^2s (1 + b\bar{b} [\beta_L\bar{\beta}_L -\SLV{\beta}_T\!\cdot\!\bar{\SLV{\beta}}_T  ]) \ , \\[1ex]
\tilde{c}_2 ={}& 2s (1  +b\bar{b}[\SLV{\beta}\!\cdot\!\bar{\SLV{\beta}]}) \ , \\[1ex]
\tilde{c}_3={}& 2\lambda_K \Big\{ (1-\beta_{\mu}^2 c^2\theta )(1 +b\bar{b} [\beta_L\bar{\beta}_L -\SLV{\beta}_T\!\cdot\!\bar{\SLV{\beta}}_T  ])
    +2s^2\theta  b\bar{b}\left[ (\SLV{\beta}_T\!\cdot\!\SLV{n}_{kT}) (\bar{\SLV{\beta}}_T\!\cdot\!\SLV{n}_{kT}) -\beta_L\bar{\beta}_L \right] 
     \nonumber \\{}& 
     -4s\theta c\theta m_{\mu}s^{-1/2} b\bar{b}\left[ (\SLV{\beta}_T\!\cdot\!\SLV{n}_{kT})\bar{\beta}_L +(\bar{\SLV{\beta}}_T\!\cdot\!\SLV{n}_{kT})\beta_L \right]\Big\} \ , \\[1ex]
\tilde{c}_{13}^R ={}& 4\beta_{\mu}\lambda_K^{1/2} m_{\mu} \Big\{
           2m_{\mu} c\theta(1 + b\bar{b} [\beta_L\bar{\beta}_L\!-\!\SLV{\beta}_T\!\cdot\!\bar{\SLV{\beta}}_T  ]) \nonumber \\ {}&
           -\!\sqrt{s}s\theta  b\bar{b}\left[ (\SLV{\beta}_T\!\cdot\!\SLV{n}_{kT})\bar{\beta}_L+(\bar{\SLV{\beta}}_T\!\cdot\!\SLV{n}_{kT})\beta_L \right]
           \Big\} \ , \\[1ex]
\tilde{c}_{13}^I ={}& 4\beta_{\mu}\lambda_K^{1/2} m_{\mu} \sqrt{s} s\theta \left[ 
            b(\SLV{\beta}_T\!\times\!\SLV{n}_{kT}) -\bar{b}(\bar{\SLV{\beta}}_T\!\times\!\SLV{n}_{kT}) \right] \ ,\\[1ex]
\tilde{c}_{12}^{R} ={}& 4\beta_{\mu} m_{\mu}s b\bar{b}(\SLV{\beta}_T \!\times\! \bar{\SLV{\beta}}_T) \ , \\[1ex]
\tilde{c}_{12}^{I} ={}& 4\beta_{\mu} m_{\mu}s (b\beta_L +\bar{b}\bar{\beta}_L ) \ , \\[1ex]
\tilde{c}_{23}^{R} ={}& 4\lambda_K^{1/2} b\bar{b} \left\{ 
               \sqrt{s}s\theta \left[ (\SLV{\beta}_T\!\times\!\SLV{n}_{kT})\!\cdot\!\bar{\beta}_L-(\bar{\SLV{\beta}}_T\!\times\!\SLV{n}_{kT})\!\cdot\!\beta_L \right]+2m_{\mu}c\theta(\SLV{\beta}_T\!\times\!\bar{\SLV{\beta}}_T)\right\} \ , \\[1ex]
\tilde{c}_{23}^{I} ={}& 4\lambda_K^{1/2} \left\{\sqrt{s}s\theta \left[ b(\SLV{\beta}_T\!\cdot\!\SLV{n}_{kT}) +\bar{b}(\bar{\SLV{\beta}}_T\!\cdot\!\SLV{n}_{kT}) \right]
               -2m_{\mu}c\theta(b\beta_L  +\bar{b}\bar{\beta}_L) 
              \right\} \ ,
\end{align}
where we have used the shorthand notation $b(x)\equiv b$ and $b(\bar{x})\equiv \bar{b}$. As expected, integration over $d\Omega d\bar{\Omega}$ results in the vanishing of all the terms involving spin correlations. Next, we make use of the identity $\int \! d\Omega/(4\pi) n(x) dx=1$, which allows one to write the total decay width as
\begin{align}
\nonumber
\label{eq:summedDW}
d\Gamma =& \frac{ds dc\theta}{64(2\pi)^3 m_{\eta^{(\prime)}}} \frac{\lambda^{1/2}_K \beta_{\mu}}{m_{\eta^{(\prime)}}^2} 2
\Big[\beta_{\mu}^2m_{\mu}^2s|F_1|^2  +s|F_2|^2  + \lambda_K(1-\beta_{\mu}^2 c\theta^2) |F_3|^2  \\
  &  +4\beta_{\mu}\lambda_K^{1/2}m_{\mu}^2c\theta \operatorname{Re}(F_1F_3^*) \Big] \ .
\end{align}
In order to quantify the $C\!P$-violating effects, one needs to construct the appropriate 
asymmetries that arise as a result of the interference of the SM \CP{}-even and the SMEFT \CP{}-odd amplitudes. Accordingly, we define the longitudinal and transverse asymmetries as follows\footnote{Note that the $C$- and $P$-odd SM contributions that may appear in the $F_{2}$ form factor are odd in $(\bar{q}\cdot k)$ and, therefore, in $\cos\theta$, which vanishes for the defined asymmetries. The same would apply to $F_4$.}

\begin{align}
\nonumber
\label{eq:AL}
A_{L} ={}& \frac{N(c\theta_{e^+}\!>0)\!-\!N(c\theta_{e^+}\!<0)}{N(c\theta_{e^+}\!>0)\!+\!N(c\theta_{e^+}\!<0)} =  \\
=& -\frac{2}{3}\frac{\int ds dc\theta \lambda^{1/2}_K\beta_{\mu} m_{\mu} \Big[ \beta_{\mu}s \operatorname{Im}F_1F_2^* +2\lambda^{1/2}_Kc\theta \operatorname{Im}F_3F_2^* \Big]}{64(2\pi)^3 m_{\eta^{(\prime)}}^3 \int d\Gamma} \ , \\ \nonumber
\label{eq:AT}
A_{T} ={}& \frac{N[s(\bar{\phi}-\phi) \!>0]\!-\!N[s(\bar{\phi}-\phi) \!<0]}{N[s(\bar{\phi}-\phi) \!>0]\!+\!N[s(\bar{\phi}-\phi) \!<0]} = \\
=& \frac{\pi}{18}\frac{\int ds dc\theta \lambda^{1/2}_K\beta_{\mu}  m_{\mu} \Big[\beta_{\mu}s \operatorname{Re}F_1F_2^* +2\lambda^{1/2}_Kc\theta \operatorname{Re}F_3F_2^* \Big]}{64(2\pi)^3 m_{\eta^{(\prime)}}^3 \int d\Gamma} \ ,
\end{align}
where the polar angles $\theta_{e^{\pm}}$ refer to those of the $e^{\pm}$ in the $\mu^{\pm}$ rest frames, $\phi(\bar{\phi})$ correspond to the azimuthal $e^{\pm}$ angles in the $\mu^{\pm}$ rest frames, and $N$ refers to the number of $\eta^{(\prime)}$ decays. It is important to highlight that only the terms associated to $\tilde{c}_{12}^{R,I}$ and $\tilde{c}_{23}^{R,I}$ contribute to the above asymmetries.
\\


\section{Results and discussion}\label{resultsdisc}

In this section, we present quantitative results for the longitudinal and transverse asymmetries by plugging the theoretical expressions for $F_{1}$ and $F_{3}$ from Ref.~\cite{Escribano:2020rfs} and the hadronic matrix elements from \cref{app:hadmatrix}, required to compute $F_2$, 
into Eqs.~(\ref{eq:AL}) and~(\ref{eq:AT}). The asymmetries for the three semileptonic processes read 
\begin{align}
 \label{eq:ALetapi0}
 A_L^{\eta\to\pi^0\mu^+\mu^-} &{} \!= -0.19(6) \operatorname{Im} c_{\ell equ}^{(1)2211} - 0.19(6) \operatorname{Im} c_{\ell edq}^{2211} - 0.020(9) \operatorname{Im} c_{\ell edq}^{2222} \ , \\[1ex] 
 A_T^{\eta\to\pi^0\mu^+\mu^-} &{} \!= 0.07(2) \operatorname{Im} c_{\ell equ}^{(1)2211} + 0.07(2) \operatorname{Im} c_{\ell edq}^{2211} + 7(3)\times 10^{-3} \operatorname{Im} c_{\ell edq}^{2222} \ , \\[1ex] 
 A_L^{\eta^\prime\to\pi^0\mu^+\mu^-} &{} \!= -0.04(8) \operatorname{Im} c_{\ell equ}^{(1)2211} - 0.04(8) \operatorname{Im} c_{\ell edq}^{2211} + 10(3)\times 10^{-3} \operatorname{Im} c_{\ell edq}^{2222}\ , \\[1ex]   
 A_T^{\eta^\prime\to\pi^0\mu^+\mu^-} &{} \!= 3(6)\times 10^{-3} \operatorname{Im} c_{\ell equ}^{(1)2211} + 3(6)\times 10^{-3} \operatorname{Im} c_{\ell edq}^{2211} - 7(2)\times 10^{-4} \operatorname{Im} c_{\ell edq}^{2222}\ , \\[1ex]  
 A_L^{\eta^\prime\to\eta\mu^+\mu^-} &{} \!= -5(39)\times 10^{-3} \operatorname{Im} c_{\ell equ}^{(1)2211} + 5(46)\times 10^{-3} \operatorname{Im} c_{\ell edq}^{2211} - 0.08(1) \operatorname{Im} c_{\ell edq}^{2222}\ , \\[1ex] 
 \label{eq:ATetaprimeeta}
 A_T^{\eta^\prime\to\eta\mu^+\mu^-} &{} \!= 7(50)\times 10^{-5} \operatorname{Im} c_{\ell equ}^{(1)2211} - 6(65)\times 10^{-5} \operatorname{Im} c_{\ell edq}^{2211} \nonumber \\
 & \ \ \ + 1(19)\times 10^{-3} \operatorname{Im} c_{\ell edq}^{2222}\ ,
\end{align}
where the error quoted accounts for both the numerical integration and the model-dependence\footnote{In particular, we use the difference between the \LCPT{} LO and NLO results as an estimation for the residual error associated to truncating the perturbative series, which in turn is used to, rather conservatively, quantify the error corresponding to the model.} uncertainties, with the latter strongly dominating over the former.

Next, in order to assess the sensitivity to new physics, one starts by estimating the expected number of events at REDTOP, which can be obtained from the projected statistics\footnote{A total production of $2.5\times 10^{13}$ $\eta/\textrm{yr}$ and $2.5\times 10^{11}$ $\eta^\prime/\textrm{yr}$ is expected \cite{test2}, with assumed reconstruction efficiencies of approximately 20\%~\cite{Sanchez-Puertas:2019qwm}.} of $5\times 10^{12}$ $\eta/\textrm{yr}$ and $5 \times 10^{10}$ $\eta^\prime/\textrm{yr}$, and the SM branching ratios for the three muonic semileptonic processes from Ref.~\cite{Escribano:2020rfs}. Accordingly, the estimated (statistical) SM backgrounds at the $1\sigma$ level, which can be assessed using $\sigma=1/\sqrt{N}$, are found to be $\sigma_{\eta\to\pi^0\mu^+\mu^-}=1.35\times 10^{-2}$, $\sigma_{\eta^\prime\to\pi^0\mu^+\mu^-}=0.105$ and $\sigma_{\eta^\prime\to\eta\mu^+\mu^-}=0.354$. It is now straightforward to estimate the REDTOP sensitivity to each of the SMEFT $C\!P$-violating Wilson coefficients from \cref{eq:operators} by setting to zero two out of the three  coefficients in Eqs.~(\ref{eq:ALetapi0}--\ref{eq:ATetaprimeeta}). The corresponding results for the three decays studied in this work are summarised in \cref{tb1}. We also show in this table the REDTOP sensitivity to the same coefficients from $\eta\to\mu^+\mu^-$ \cite{Sanchez-Puertas:2018tnp}, as well as the bounds set by the nEDM experiment 
using the most recent measurements from Ref.~\cite{nEDM:2020crw} (the bounds derived from $D_s^-\to\mu\bar{\nu}_\mu$ decays are weaker and, thus, we do not quote them \cite{Sanchez-Puertas:2019qwm}). 
It must be highlighted that, strictly speaking, the nEDM experiment sets bounds on a particular linear combination of the three Wilson coefficients, which raises the question about possible cancellations that may weaken the nEDM bounds. From Eqs.~(4.17) and (4.20) in Ref.~\cite{Sanchez-Puertas:2018tnp}, one can clearly see that partial cancellations are possible for $c_{\ell equ}^{(1)2211} \sim c_{\ell edq}^{2211}$, which would weaken the nEDM bounds by an order of magnitude.\footnote{More drastic cancellations would require what it seems to us a large degree of fine-tuning. Furthermore, it seems unlikely that these cancellations would remain stable at higher order corrections.} Even in such scenario, REDTOP would still not be competitive. 

Clearly, the most competitive observable amongst those studied in this work is the longitudinal asymmetry of the $\eta\to\pi^0\mu^+\mu^-$ decay. As well as this, it can be seen that the constraints imposed by the $\eta^\prime$ semileptonic decays are comparatively much weaker, which is down to the $\eta^\prime$ REDTOP projected statistics being two orders of magnitude smaller than that of the $\eta$. If one compares the sensitivities obtained from the $\eta^{(\prime)}\to\pi^0\mu^+\mu^-$ and $\eta^\prime\to\eta\mu^+\mu^-$ decays to the \CP{}-violating Wilson coefficients with the bounds extracted from nEDM experiments, one must conclude that the projected REDTOP statistics are not competitive enough for the above semileptonic processes, which can be attributed to the isospin-breaking suppression in the hadronic matrix elements, subject to the assumption that new physics can be parametrised by the SMEFT. Consequently, the leptonic $\eta\to\mu^+\mu^-$ decay studied in Ref.~\cite{Sanchez-Puertas:2018tnp} remains the most promising channel to be studied at REDTOP. 
\\

\begin{table}[t]
\centering
\captionsetup{font=small}
\caption{
Summary of REDTOP sensitivities to (the imaginary parts of) the Wilson coefficients associated to the SMEFT $C\!P$-violating operators in \cref{eq:operators} for the processes studied in this work, as well as the $\eta\to\mu^+\mu^-$ decay analysed in Ref.~\cite{Sanchez-Puertas:2018tnp}. In addition, the upper bounds from nEDM experiments are given in the last row for comparison purposes.}
{\def\arraystretch{1.2}\tabcolsep=45pt
\small 
\begin{tabular}[c]{@{\hskip 0.1in}c @{\hskip 0.2in}c @{\hskip 0.2in}c @{\hskip 0.2in}c @{\hskip 0.2in}c @{\hskip 0.1in}c}
 \hline \hline \\[-12pt]
Process & Asymmetry & $ \operatorname{Im} c_{\ell equ}^{(1)2211}$ & $ \operatorname{Im} c_{\ell edq}^{2211} $ & $\operatorname{Im} c_{\ell edq}^{2222} $ \\ \\[-12pt]
\hline \hline
\multirow{2}{*}{$\eta\to\pi^0\mu^+\mu^-$} & $A_L$ & $0.0695$ & $0.0720$ & $0.686$ \\ 
& $A_T$ & $0.194$ & $0.203$ & $1.93$  \\
\hline
\multirow{2}{*}{$\eta^\prime\to\pi^0\mu^+\mu^-$} & $A_L$ & $2.36$ & $2.56$ & $10.96$ \\ 
& $A_T$ & $33.1$ & $35.8$ & $154$  \\
\hline
\multirow{2}{*}{$\eta^\prime\to\eta\mu^+\mu^-$} & $A_L$ & $67.5$ & $78.5$ & $4.46$ \\ 
& $A_T$ & $5264$ & $5549$ & $328$  \\
\hline \hline
$\eta\to\mu^+\mu^-$ & $A_L$ & $0.007$ & $0.007$ & $0.005$ \\
\hline \hline
nEDM & - & $\leq 0.001$ & $\leq 0.002$ & $\leq 0.02$ \\
\hline \hline
\end{tabular}
}
\label{tb1}
\end{table}


\section{Conclusions}\label{conclusions}

In this work, we have analysed in detail possible effects of physics BSM via $C\!P$ violation in $\eta^{(\prime)}\to\pi^0\mu^+\mu^-$ and $\eta^\prime\to\eta\mu^+\mu^-$ decays. This is particularly timely at present as the REDTOP experiment is studying the possibility of using polarisation techniques to study $C\!P$-violating new-physics effects. Assuming that BSM $C\!P$-violation appears in Nature via new heavy degrees of freedom, the use of the SMEFT is justified, which in turn provides a convenient connection to different observables, such as those from nEDM experiments and $D_s^-\to\mu\bar{\nu}_\mu$ decays. The outcome of the present work is that the predicted statistics at REDTOP will fall short to detect any $C\!P$-violating effects in the semileptonic $\eta^{(\prime)}\to\pi^0\mu^+\mu^-$ and $\eta^\prime\to\eta\mu^+\mu^-$ decays, should one take into account the constraints set by nEDM and $D_s^-\to\mu\bar{\nu}_\mu$. This stands in stark contrast with the $\eta\to\mu^+\mu^-$ decay studied in Ref.~\cite{Sanchez-Puertas:2018tnp} and can be understood by the fact that the less constrained strange quark contribution (cf.~\cref{tb1}) is of isospin-breaking origin, which is very small in Nature.
Accordingly, the leptonic $\eta\to\mu^+\mu^-$ decay is still the most promising channel to be investigated at REDTOP in search of new-physics signatures via $C\!P$-violating effects using muon polarimetry.
\\


\section*{Acknowledgements}

We would like to thank Corrado Gatto for the useful comments about the REDTOP experiment and its setup.
This work is supported by the Secretaria d'Universitats i Recerca del Departament d'Empresa i Coneixement de la Generalitat de Catalunya under the grant 2017SGR1069, the Ministerio de Econom\'{i}a, Industria y Competitividad (grant FPA2017-86989-P and SEV-2016-0588), the Ministerio de Ciencia e Innovaci\'on (grant PID2020-112965GB-I00), and the European Union’s Horizon 2020 Research and Innovation Programme (grants no.~754510 (EU, H2020-MSCA-COFUND2016) and no.~824093 (H2020-INFRAIA-2018-1)).~IFAE is partially funded by the CERCA program of the Generalitat de Catalunya.
\\


\appendix

\section{Kinematics and phase space conventions}\label{app:kin}

\begin{figure}\centering
\includegraphics[width=0.4\textwidth]{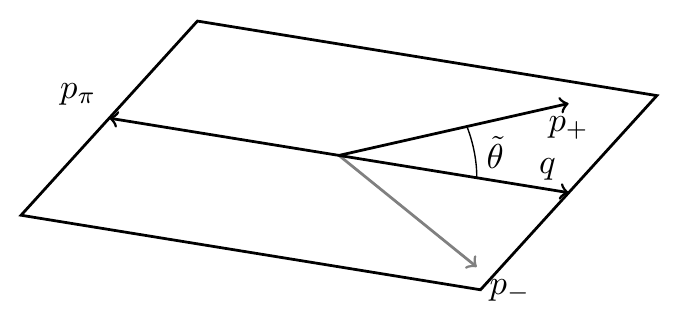}
\includegraphics[width=0.3\textwidth]{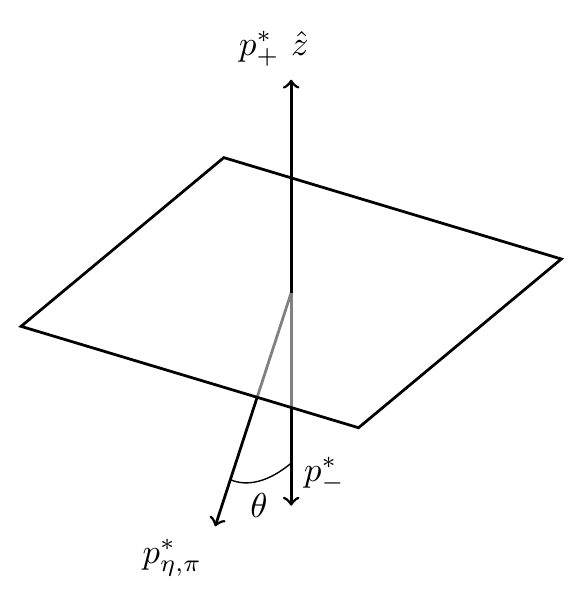}
\caption{Left: the coordinates in the $\eta$ rest frame. Right: the coordinates in the dilepton frame. Note that the longitudinal ($\hat{z}$) axis is chosen along the $\mu^+$ direction and that $\theta\neq \tilde{\theta}$. Only $\vec{p}_{\pm}\sin\tilde{\theta} = \vec{p}^{\ *}_{\pm}\sin\theta$ is preserved in the boost.\label{fig:refframe}}
\end{figure}

In this work, the phase space is described in terms of invariant masses and the $\mu^+$ angle in the dilepton rest-frame, as shown in \cref{fig:refframe}. 
This choice is convenient for the computation of the scalar products involving spin directions. The independent momenta for the $\eta\to\pi^0\mu^+\mu^-$ decay can be written as 
\begin{equation}
q = p_{\mu^+} + p_{\mu^-} = p_{\eta} -p_{\pi} \ , \qquad \bar{q} = p_{\mu^+} - p_{\mu^-} \ , \qquad k = p_{\eta}+p_{\pi} \ .
\end{equation}
The relevant scalar products can, in turn, be expressed as 
\begin{equation}
 q^2 = s \ , \ \ \bar{q}^2 = 4m_{\mu}^2 -s \ , \ \ k^2 = 2(m_{\eta}^2 +m_{\pi}^2) -s \ , \ \ q\cdot\bar{q} =0 \ , 
\end{equation}
\begin{equation}
q\cdot k = m_{\eta}^2 -m_{\pi}^2 \ , \ \  \bar{q}\cdot k = \beta_{\mu} \lambda^{1/2}_Kc\theta \ ,
\end{equation}
with $\beta_{\mu}^2= 1-4m_{\mu}^2/s$ and $\lambda_K\equiv\lambda(m_{\eta}^2,m_{\pi}^2,s)$, where $\lambda(a,b,c)=a^2+b^2+c^2-2ab-2ac-2bc$. Across the entire manuscript, we use $c\theta\equiv \cos\theta$ and $s\theta \equiv \sin\theta$ for economy of notation. With these conventions, the differential decay width is 
\begin{equation}
d\Gamma = \frac{1}{64(2\pi)^3 m_{\eta}} \frac{\lambda^{1/2}_K \beta_{\mu}}{m_{\eta}^2}  |\mathcal{M}|^2 \  ds dc\theta \ .
\end{equation}
It is also useful to quote all $4$-momenta in the dilepton rest frame
\begin{equation}
q^* = (\sqrt{s},0,0,0)\ ,  \quad \bar{q}^* = (0,0,0,\sqrt{s}\beta_{\mu})\ , \quad p^*_{\pm} = \sqrt{s}/2(1,0,0,\pm\beta_{\mu}) \ ,  
\end{equation}
\begin{equation}
n^* = (+\gamma\beta_{\mu}n_L, \SLV{n}_T, \gamma n_L)\ , \quad \bar{n}^* = (-\gamma\beta_{\mu}\bar{n}_L, \bar{\SLV{n}}_T, \gamma \bar n_L)\ ,  
\end{equation}
\begin{equation}
k^* = \frac{1}{\sqrt{s}}(m_{\eta}^2 -m_{\pi}^2, \lambda^{1/2}_K s\theta  \SLV{n}_{kT},  -\lambda^{1/2}_K c\theta)\ , 
\end{equation}
\begin{equation}
   p^*_{\eta(\pi)} = \frac{1}{2\sqrt{s}}(m_{\eta}^2 \pm s -m_{\pi}^2, \lambda^{1/2}_K s\theta  \SLV{n}_{kT}\ ,  -\lambda^{1/2}_K c\theta)\ ,
\end{equation}where $\SLV{n}_T(\bar{\SLV{n}}_T)$ and $n_L(\bar{n}_L)$ are, respectively, the transverse and longitudinal $\mu^{\pm}$ spin components with respect to the $\mu^+$ direction, and $\SLV{n}_{kT}$ is a unit vector representing the $k$ momentum transverse to the $\mu^+$ direction. Note that $\SLV{n}_T$, $\bar{\SLV{n}}_T$ and $\SLV{n}_{kT}$ are 2-dimensional objects. The corresponding expressions for the other two processes are found by substituting $\eta \to \eta^\prime$ for $\eta^\prime \to \pi^0 \mu^+\mu^-$, and $\eta \to \eta^\prime$ and $\pi^0 \to \eta$ for $\eta^\prime \to \eta \mu^+\mu^-$.

The spin projectors, required when one does not sum over spins, are given here for completeness%
\begin{align}
 u(p,\bar{\lambda} \bar{\SLV{n}})\bar{u}(p,\bar{\lambda} \bar{\SLV{n}})&=\frac{1}{2}(\slashed{p}+m)(1+\bar{\lambda} \gamma^5\slashed{\bar{n}}) \ , \\ 
 v(p,\lambda \SLV{n})\bar{v}(p,\lambda \SLV{n})&=\frac{1}{2}(\slashed{p}-m)(1+\lambda \gamma^5\slashed{n}) \ .
\end{align} 
Note that the results obtained from the above equations are easily adapted to the Bouchiat-Michel formulae, Refs.~\cite{Bouchiat:1958yui,Michel:1959dvg}, required in the spin-density formalism (see Ref.~\cite{Haber:1994pe}, Sect.~1.6). 

\section{Polarised amplitudes}\label{app:polamps}

The $c_i$ coefficients in \cref{eq:masterf} are used as an intermediate step in our calculation but are relevant for implementing the decay processes in a Monte Carlo program, where the subsequent polarised $\mu^{\pm}$ decays are taken care of 
by {\textsc{Geant4}}. On these grounds, we provide them here for completeness
\begin{align}
c_1 ={}& 2\beta_{\mu}^2 m_{\mu}^2s (1 +\SLV{n}_T\!\cdot\!\bar{\SLV{n}}_T -n_L\bar{n}_L) \ , \\[1ex]
c_2 ={}& 2s (1 -\SLV{n}\!\cdot\!\bar{\SLV{n}}) \ , \\[1ex]
c_3={}& 2\lambda_K \Big\{ (1-\beta_{\mu}^2 c^2\theta )(1 +\SLV{n}_T\!\cdot\!\bar{\SLV{n}}_T -n_Ln_L)
    +2s^2\theta \left[n_L\bar{n}_L - (\SLV{n}_T\!\cdot\!\SLV{n}_{kT}) (\bar{\SLV{n}}_T\!\cdot\!\SLV{n}_{kT}) \right] 
     \nonumber \\ {}& 
     +4s\theta c\theta m_{\mu}s^{-1/2}\left[ (\SLV{n}_T\!\cdot\!\SLV{n}_{kT})\bar{n}_L +(\bar{\SLV{n}}_T\!\cdot\!\SLV{n}_{kT})n_L \right]
     \Big\} \ , \\[1ex]
c_{13}^R ={}& 4\beta_{\mu}\lambda_K^{1/2} m_{\mu} \Big\{
           2m_{\mu} c\theta(1 +\SLV{n}_T\!\cdot\!\bar{\SLV{n}}_T -n_L\bar{n}_L) \nonumber \\
  &+\sqrt{s}s\theta \left[ (\SLV{n}_T\!\cdot\!\SLV{n}_{kT})\bar{n}_L +(\bar{\SLV{n}}_T\!\cdot\!\SLV{n}_{kT})n_L \right]\Big\} \ , \\[1ex]
c_{13}^I ={}& -4\beta_{\mu}\lambda_K^{1/2} m \sqrt{s} s\theta \left[ 
            (\SLV{n}_T\!\times\!\SLV{n}_{kT}) +(\bar{\SLV{n}}_T\!\times\!\SLV{n}_{kT}) \right]\ ,\\[1ex]
c_{12}^{R} ={}& -4\beta_{\mu} m_{\mu}s (\SLV{n}_T \!\times\! \bar{\SLV{n}}_T)\ , \\[1ex]
c_{12}^{I} ={}& 4\beta_{\mu} m_{\mu}s (\bar{n}_L -n_L)\ , \\[1ex]
c_{23}^{R} ={}& -4\lambda_K^{1/2} \left\{\sqrt{s}s\theta \left[ (\SLV{n}_T\!\times\!\SLV{n}_{kT})\!\cdot\!\bar{n}_L-(\bar{\SLV{n}}_T\!\times\!\SLV{n}_{kT})\!\cdot\!n_L \right]+2m_{\mu}c\theta (\SLV{n}_T\!\times\!\bar{\SLV{n}}_T)\right\}\ , \\[1ex]
c_{23}^{I} ={}& -4\lambda_K^{1/2} \left\{2m_{\mu}c\theta(\bar{n}_L -n_L)+\sqrt{s}s\theta \left[ (\SLV{n}_T\!\cdot\!\SLV{n}_{kT}) -(\bar{\SLV{n}}_T\!\cdot\!\SLV{n}_{kT}) \right]\right\}\ ,
\end{align}
where the shorthand notation $\lambda\SLV{n}\to\SLV{n}$ and $\bar{\lambda}\bar{\SLV{n}}\to\bar{\SLV{n}}$ has been employed.

\section{Polarised muon decay}\label{app:polmudec}

In order to study the relevant asymmetries, it is necessary to supplement the $\eta^{(\prime)}\to\pi^0\mu^+\mu^-$ and $\eta^{\prime}\to\eta\mu^+\mu^-$ processes with the subsequent $\mu^{\pm}$ decays. The corresponding result reads~\cite{Sanchez-Puertas:2018tnp}
  \begin{equation}
  \label{eq45}
    \big\vert\mathcal{M}\left(\mu^{+},\lambda \SLV{n}\right)\big\vert^2 = 64G_F^2 k_{\alpha}( p_{\beta} + \lambda  m_{\mu}n_{\beta} )q_1^{\alpha}q_2^{\beta}\ .
  \end{equation}
Including the phase space and integrating over the neutrino spectra (note that the muon rest frame is employed), the above result becomes\footnote{In the second line, the result of integrating over $d\Omega dx$ has been employed, which introduces $\epsilon=m_e^2\left[m_e^2(m_\mu^2-m_e^2)^2 +6m_{\mu}^6 +2m_e^2m_{\mu}^4\left(1 +6\ln(m_e/m_{\mu}) \right)\right]/(m_e^2 +m_{\mu}^2)^{4}$.}$^,$\footnote{Note that \cref{eq:mudecayexact} is the SM result from Ref.~\cite{Zyla:2020zbs}, as well as the expression implemented in {\textsc{Geant4}}, though this simulation package includes, in addition, radiative corrections.}
   \begin{align}
      \frac{d\Gamma(\mu^{+},\lambda \SLV{n})}{dxd\Omega} &{}= 
      \frac{m_{\mu}}{24\pi^4}W_{e\mu}^4G_F^2\beta x^2 n(x,x_0)\left[  
                1 - \lambda b(x,x_0) \SLV{\beta}\cdot\SLV{n}  
      \right] ,\\\label{eq:mudecayexact}
      d\textrm{BR}(\mu^{+},\lambda \SLV{n}) &{}= 
      \frac{d\Omega}{4\pi} \frac{2x^2\beta}{1-2\epsilon}   n(x,x_0)\left[  
                1 - \lambda b(x,x_0) \SLV{\beta}\cdot\SLV{n}  
      \right]dx\ ,
   \end{align}
with $n(x,x_0)=(3-2x-x_0^2/x)$ and $n(x,x_0)b(x,x_0)=2- 2x -\sqrt{1-x_0^2}$. 
Moreover, $W_{e\mu}=(m_{\mu}^2+m_e^2)/2m_{\mu}$ is the maximum positron energy, $x=E_e/W_{e\mu}$ is the reduced positron energy, $x_0=m_e/W_{e\mu}$ is the minimum reduced positron energy, and $\beta=\sqrt{1-x_0^2/x^2}$. Typically, the approximation $m_e/m_{\mu}\to 0$ is employed, which results in the simpler expression
\begin{equation}\label{eq:mudecayapp}
      d\textrm{BR}(\mu^{+},\lambda \SLV{n})= 
      \frac{d\Omega}{4\pi} n(x)  \left[  
                1 - \lambda b(x) \SLV{\beta}\cdot\SLV{n}  
      \right]dx \ ,
\end{equation}
with $x=2E_e/m_{\mu}$, $n(x)=2x^2(3-2x)$ and $b(x)=(1-2x)/(3-2x)$. The corresponding expressions for the $\mu^-$ are found by replacing $n\to - \bar{n}$ or $\SLV{n}\to - \bar{\SLV{n}}$ on the right hand side of Eqs.~(\ref{eq45}--\ref{eq:mudecayapp}).
\\

\bibliographystyle{apsrev4-1}
\bibliography{refs}

\end{document}